\documentclass[aps,pre,twocolumn,float]{revtex4}
\usepackage{amsmath,bm,epsfig}

\def\Fbox#1{\vskip1ex\hbox to 8.5cm{\hfil\fboxsep0.3cm\fbox{%
  \parbox{8.0cm}{#1}}\hfil}\vskip1ex\noindent}  

\newcommand{\B}[1]{{\bm{#1}}}
\let \= \equiv \let\*\cdot \let\~\widetilde \let\-\overline

\begin{document}
\title{Relaxation Mechanisms in Glassy Dynamics: the Arrhenius and Fragile Regimes}
\author{H. George E. Hentschel$^{1}$, Smarajit Karmakar$^{2}$, Itamar Procaccia$^{1}$ and Jacques Zylberg$^{1}$}
\affiliation{$^{1}$ Department of Chemical Physics, The Weizmann
 Institute of Science, Rehovot 76100, Israel\\
 $^{2}$ Departimento di Fisica, Universit\'a di Roma ``La Sapienza '', Piazzale Aldo Moro 2, 00185, Roma, Italy}
\date{\today}
\begin{abstract}
Generic glass formers exhibit at least two characteristic changes in their relaxation behavior, first to an Arrhenius-type relaxation at some characteristic temperature, and then at a lower characteristic
temperature to a super-Arrhenius (fragile) behavior. We address these transitions
by studying the statistics of free energy barriers for different systems at different temperatures and space dimensions. We present a clear evidence for changes in the dynamical behavior at the transition to Arrhenius and then to a super-Arrhenius behavior. A simple model is presented, based on the idea of competition between single-particle and cooperative dynamics. We argue that Arrhenius behavior can take place as long as there is enough free volume for the completion of a simple $T1$ relaxation process. Once free volume is absent one needs a cooperative mechanism to `collect' enough free volume. We show that this model captures all the qualitative behavior observed in simulations throughout the considered temperature range.
\end{abstract}
\maketitle

\begin{figure}
\includegraphics[scale = 0.6]{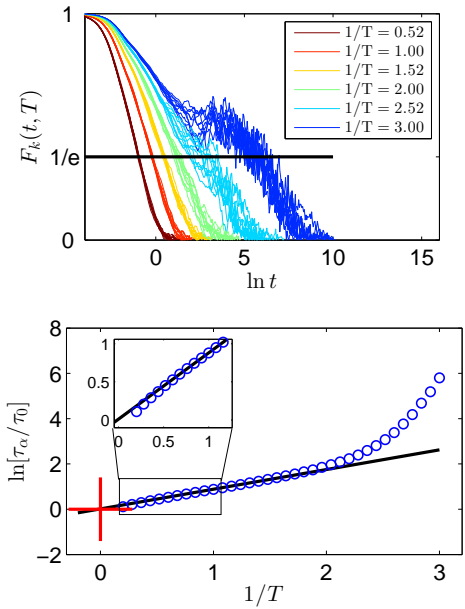}
\caption{Color online. Upper panel: the time dependence of ten
 realizations of the unaveraged intermediate scattering
function Eq. (\ref{defFk}), for a range of temperatures (decreasing from left to
right). By definition the relaxation time $\tau_\alpha$ is the time at which the function reaches
the value of $1/e$. Notice the increase in the dispersion in $\tau_\alpha$ when temperature is lowered.  Lower panel: the quenched average relaxation time
$\langle \tau \rangle$ in a log-lin plot vs $1/T$, compared to an Arrhenius
temperature dependence. Inset: the pre-Arrhenius regime at high temperatures.
The simulation for this figure were done using the binary Lennard-Jones model whose inter-particle potential can
be found for example in \cite{11KLPZ}}
\label{relaxation}
\end{figure}
\section{introduction}
A full understanding of the complex dynamical scenarios accompanying the glass transition requires a
resolution of the statistics of free energy barriers at any given temperature. Such information is very
hard to come by. Trying to measure such free energy barriers using the energy landscape of a typical
glass former is a daunting procedure that has never been achieved. In this paper we propose a method
that provides us with a decent approximation of the statistics of the free energy barriers. We do this
by measuring ``quenched" rather than ``annealed" relaxation times.
The usual procedure for extracting a relaxation time from numerical simulations of any glassy
system is the ``annealed" procedure. One takes many realizations of a super-cooled liquid
at a given temperature, measures a typical relaxation function for each realization, averages
the relaxation functions over the ensemble and finally extracts a relaxation time $\tau_\alpha(T)$
from this average relaxation function by, say, determining when the average function reaches $1/e$ of its
initial value \cite{09Cav}. The subscript $\alpha$ designates the `main' slow relaxation mechanism in glassy dynamics, and is the only one that is typically resolved in numerical simulations. This procedure is automatically performed in experiments where the macroscopic nature of the relaxing systems provides self-averaging resulting in `smooth' relaxation functions that hardly fluctuate from realization to realization. In this paper we propose that significant simulational insight on glassy relaxation can be gained by adopting a ``quenched" procedure. In this procedure a relaxation time $\tau_\alpha$ is extracted from each
and every realization by determining when the appropriate relaxation function reaches a value of $1/e$, cf. Fig. \ref{relaxation} upper panel. Finally an average is taken over the ensemble to provide $\langle \tau_\alpha \rangle(T)$. A typical such
quenched relaxation time for a Lennard-Jones binary glass is shown in the lower panel of Fig. \ref{relaxation}. Here we measured the self part of the intermediate scattering function for each realization separately (upper panel), i.e. the function
\begin{equation}
F_k(t;T) \equiv  \case{2}{N}\sum_{i=1}^{N/2} \exp\left\{{i\B k \cdot [\B r_i(t)-\B r_i(0)]}\right\}  \ . \label{defFk}
\end{equation}
where the index $i$ runs over half the particles in the binary mixture (those with longer interaction length; we assume the mass of all the particles to be the same, $m=1$). One observes
the usual dramatic slowing down, such that  $\langle \tau_\alpha \rangle(T)$ grows rapidly when $T$ decreases, first in a pre-Arrhenius form (see inset), then in an Arrhenius form, linear in $1/T$, and later in a faster, super-Arrhenius form, which is
referred to as `fragile' behavior in the glass community \cite{88Ang}. In this paper we will be interested predominantly in the transitions to and from the Arrehenius regime, for which the quenched procedure is
particularly illuminating.

The advantage of the quenched over the annealed procedure is that it allows a particularly transparent
treatment of the statistics of free energy barriers. Having an ensemble of $\tau_\alpha$ values for a given ensemble of super-cooled systems, we can define a typical free energy scale by writing for each $\tau_\alpha$ value an equation
\begin{equation}
\tau_\alpha \equiv \tau_0 \exp{(F/T)} \ , \label{defE}
\end{equation}
where we choose units such that Bolzmann's constant equals unity. Inverting Eq. (\ref{defE}), every value of $\tau_\alpha$ yields a value of $F$. The free energy has in principle a contribution from an energy barrier $E$ and an entropic contribution stemming from a degeneracy $g$,
\begin{equation}
F = E -T\ln g \ .
\end{equation}

Of course, this definition makes sense when the dynamics involves escaping an energy barrier.  The prefactor $\tau_0$ is by definition
the relaxation time inside the confined state at every temperature $T$, and we assume that it is fixed for all the realizations in the ensemble:
\begin{equation}
\tau_0 = C/\sqrt{T} \ . \label{deftau0}
\end{equation}
The dependence on $1/\sqrt{T}$ stems simply from the estimating $v$, the typical particle velocity, from equipartition as $v^2=T$ and $\tau_0$ is then
a typical length-scale (a cage magnitude) over $v$. For our purpose
the constant $C$ is determined by requiring that the plot of $\tau_\alpha/\tau_0\to 1$ as $T\to \infty$, see Fig. \ref{relaxation} (red cross) and is $\mathcal{O}(1)$.
\begin{figure}
\hskip -20pt
\includegraphics[scale = 0.6]{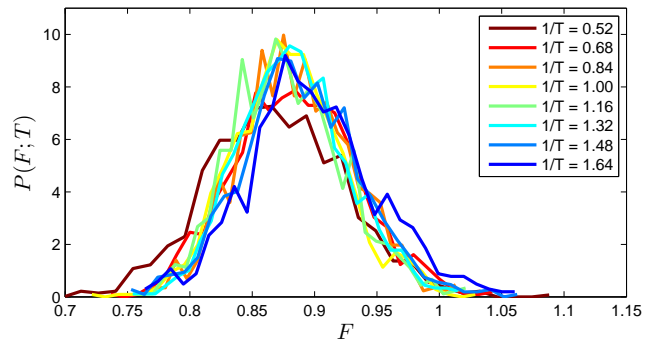}
\caption{Color online: Distribution functions $P(F;T)$ in the Arrhenius regime of the Lennard Jones system in dimension $d=2$, shown
for a range of temperatures. Neither the peak nor the width of the approximately Gaussian distribution depends on temperature in this regime, meaning for this relaxation process $g=1$.}
\label{gaussArr}
\end{figure}

\section{The Arrhenius regime}
Our first observation is that in
the Arrhenius regime the distribution of free energies $F$ is always Gaussian, and is always very sharply peaked. Since $F$ in the Arrhenius regime is temperature independent, we conclude that $g=1$, and $F=E$. As an example we show again the binary Lennard-Jones system for which we have used 1000 realizations for each value of $T$ to generate the distributions of $F$ shown in Fig. \ref{gaussArr}. We conclude that in the Arrhenius regime to a very good approximation
\begin{equation}
P(F) = \frac{1}{\sqrt{2\pi\sigma^2}}e^{\frac{-(F-\langle F\rangle )^2}{2\sigma^2}} \ ,
\end{equation}
with the peak value $\langle F\rangle$ {\em and} the width $\sigma$ {\em unchanged} throughout the Arrhenius regime.
In the present regime we denote $\langle F\rangle\equiv E_a$.
It is noteworthy that the distributions are also very sharply peaked; to see this note that the
quenched average can be now computed from the Gaussian integral as
\begin{equation}
\langle \tau_\alpha \rangle(T) = \tau_0 \exp[\frac{E_a}{T} +\frac{\sigma^2}{T^2}] \ . \label{quenched}
\end{equation}
Alternatively,
\begin{equation}
\ln\frac{\langle \tau_\alpha \rangle(T)}{\tau_0} = \frac{E_a}{T} +\frac{\sigma^2}{T^2} \ . \label{simple}
\end{equation}

\begin{figure}
\includegraphics[scale = 0.6]{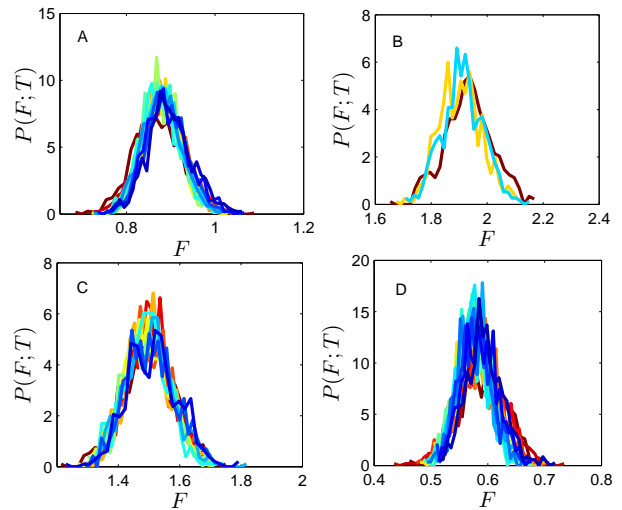}
\caption{Color online: Distributions $P(F;T)$ for the purely repulsive system (A), the Lennard-Jones model in $d=3$ (B), the Kob-Andersen model (C) and the polydisperse model (D) in the Arrhenius regime. The distributions $P(F;T)$ for different temperatures overlap throughout the Arrhenius regime.}
\label{examplesArrhenius}
\end{figure}

For the data shown in Fig. \ref{gaussArr} $E_a=\mathcal{O}(1)$ and $\sigma\approx 0.05$. Thus for the Arrhenius range of temperatures $T=\mathcal{O}(1)$ the contribution of the width $\sigma^2\approx 2.5\times 10^{-3}$ is of the order of $0.1\%$ in Eq. (\ref{quenched}). Obviously, for larger systems we expect $\sigma$ to be even smaller, tending to zero for $N\to \infty$. It appears that all the relaxation events contributing to the Arrhenius regime belong to a tight group of similar events with a very well defined energy barrier $E_a$ and a negligible dispersion. In fact, we find that this conclusion is not particular either to the Lennard-Jones system or to $d=2$. In Fig. \ref{examplesArrhenius} we show similar results for other glass formers with A) purely repulsive potential \cite{09LPZ} at $d=2$, B) the Lennard-Jones system \cite{11KLPZ} at $d=3$, C) the Kob-Andersen model \cite{93KA} at $d=2$ and D) the polydisperse model \cite{08WSB} at $d=2$. The physical interpretation of these observations is discussed below, cf. Sect. \ref{modelsec}.

\section{The pre-Arrhenius regime}
At high temperatures we observe (see for example Fig. \ref{fullT-Lennard}) that $P(F)$ does not have
a fixed dispersion (it widens when $T$ increases) and its peak moves to
lower free energy values. This observation is generic for all the models studied here. The physical
interpretation of this observation is again discussed in Sect. \ref{modelsec}.

\section{Transition to super-Arrhenius}
\begin{figure}
\hskip -15pt
\includegraphics[scale = 0.6]{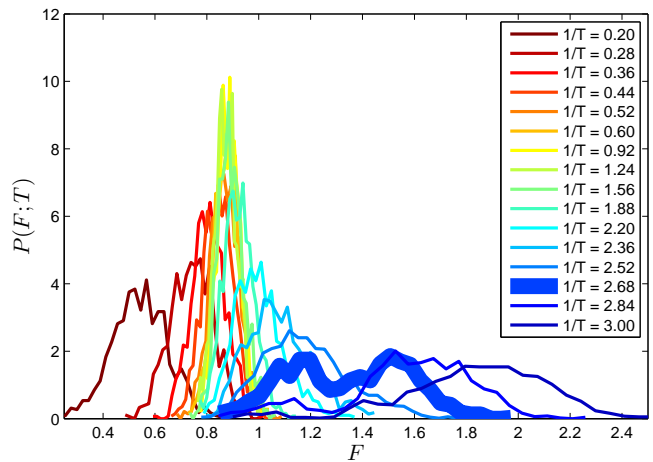}
\caption{Color online: Distributions $P(F;T)$ for the Lennard-Jones system at $d=2$ with $N=6400$ for the whole available temperature range.}
\label{fullT-Lennard}
\end{figure}

\begin{figure}
\hskip 30pt
\includegraphics[scale = 0.65]{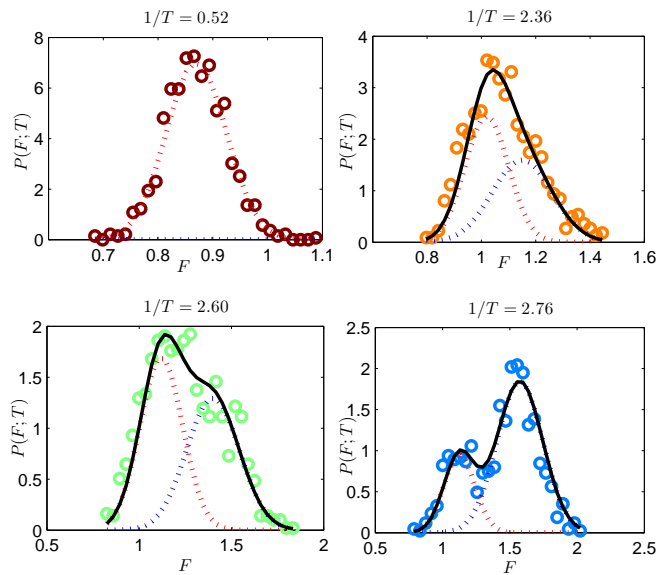}
\caption{Color online: The transition in relaxation mechanism shown in Fig. \ref{fullT-Lennard} as seen by the replacement of one Gaussian distribution $P(F)$ by another.}
\label{transition}
\end{figure}

Obviously, one might expect that when $T$ reaches a value of $T\approx Const\times E_a$ something interesting should happen. The physical relaxation mechanism whose energy barrier is $E_a$ should run out of steam, and other mechanisms, if they exist, should emerge. This phenomenon is beautifully seen in the Lennard-Jones system, see Fig. \ref{fullT-Lennard}. Indeed, at $1/T\approx 1.5$ the Arrhenius process begins to disappear and we observe at first a gradual increase in both $\langle F\rangle$ (the peak position) and its width $\sigma$. In the system for which $N=6400$, we find that at $1/T\approx2.7$ the high temperature Gaussian peak disappears in favor of a second Gaussian peak which appears to replace it. We will denote by $T^*$ the temperature at which both process coexist with equal probability. The Gaussian of the second process marches to the right with increasing both of $\langle F\rangle$ and $\sigma$. The transition region is depicted in Fig. \ref{fullT-Lennard} and is shown in more detail in Fig. \ref{transition}. For the systems of the size studied in these simulations, by the time that the second relaxation mechanism takes over, the contribution of the variance $\sigma$ to the quenched average relaxation time Eq. (\ref{quenched}) is no longer negligible. Thus for our example at $T=1/3$ $\langle F \rangle\approx 2$ and $\sigma\approx 0.25$ such that $\sigma^2 /\langle F\rangle T\approx 0.1$. At lower temperature the dispersion effect will contribute more and more to the relaxation time. Of course, also here we expect $\sigma$ to decrease to zero when $N\to \infty$. It is useful however to keep the $\sigma$ dependence for the simulations at hand.

\begin{figure}
\hskip -25pt
\includegraphics[scale = 0.6]{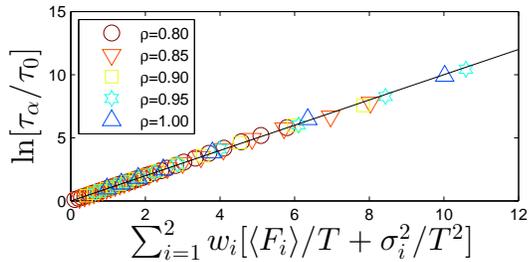}
\caption{Color online: A test of Eq.(\ref{simple}). Shown are data for the binary Lennard-Jones model at d=2 and at
different densities.}
\label{fig7}
\end{figure}

We should note that Eq. (\ref{simple}) remains valid throughout the temperature range. We can rewrite it, denoting $x_i\equiv \langle F_i\rangle/T$ and introducing $w_i$ for the weighted sum of the two different mechanisms in the form
\begin{equation}
\ln\frac{\langle \tau_\alpha \rangle(T)}{\tau_0} = \sum^2_{i=1} w_i \left[ x_i+\frac{\sigma_i^2}{\langle F_i\rangle^2} x_i^2\right] \ .
\end{equation}
We test this formula for the data corresponding to the distributions found in Fig. \ref{fullT-Lennard} and demonstrate the perfect agreement in Fig. \ref{fig7}.
Note that the relaxation time varies in this range of temperatures by close to 5 orders of magnitude.
\begin{figure}
\hskip -20pt
\includegraphics[scale = 0.8]{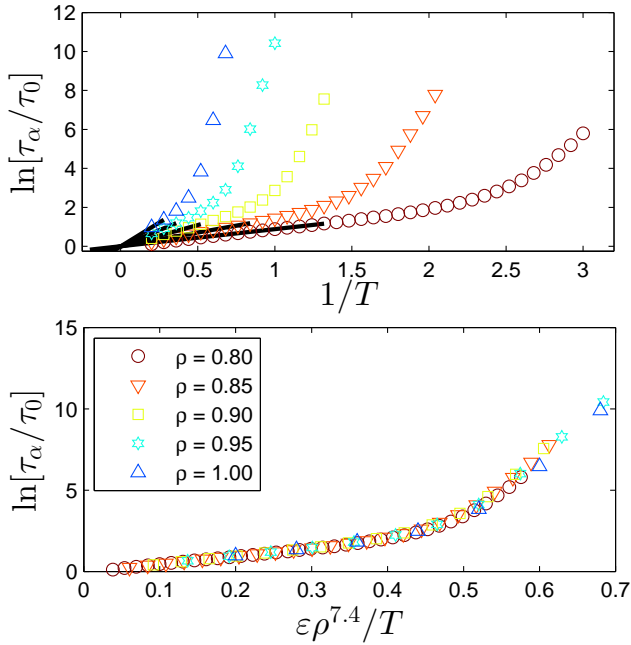}
\caption{Color online. Upper panel: Usual plots of the non-dimensional relaxation time in logarithmic scale vs. $1/T$ for
five different values of the density, $\rho=0.80, 0.85, 0.90, 0.95, 1.00$.
Lower panel: The very same data collapsed by non-dimensionalizing the abscissa by the energy scale $\varepsilon\rho^x$.}
\label{fig8}
\end{figure}

\section{Scaling}

\begin{figure}
\hskip -7pt
\includegraphics[scale = 0.6]{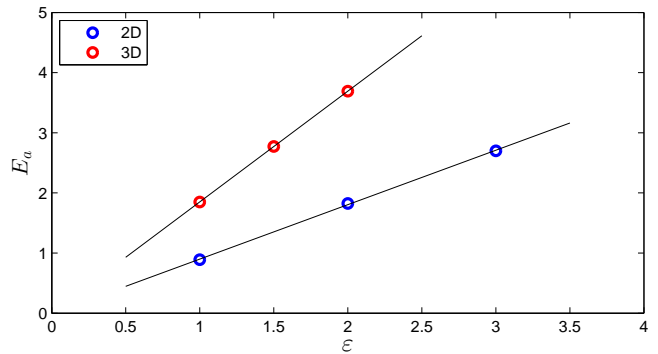}
\caption{Color online: Demonstration of the linear dependence of the mean Arrhenius energy $E_a$ on the
depth of the potential $\varepsilon$.}
\label{linear}
\end{figure}

In Ref. \cite{ludovic} it was shown that relaxation times of models of the type discussed above exhibit density scaling. For different densities $\rho$ plots of $\ln\langle \tau_\alpha\rangle/\tau_0$ vs. $1/T$ fall on different curves \cite{02AKT,04TKMA}. Once plotted as a function of $\varepsilon\rho^x/T$ with an appropriate value of $x$ the data collapses onto one curve. Density scaling is easiest to justify when one uses a purely repulsive potential of the form $\phi(r)\sim \varepsilon r^{-y}$ where $\varepsilon$ sets the energy scale of the interparticle potential and $r$ is measured in dimensionless units. Then the only relevant length scale $\xi$ in the system is determined by the density as $\xi\sim \rho^{-1/d}$ where $\rho$ is the density in dimensionless units. Obviously, when we plot $\ln\langle \tau\rangle/\tau_0$ vs. $1/T$ (see upper panel of Fig. \ref{fig8}) we present a dimensionless number in terms of a dimensional quantity which is inappropriate. To correct this, we need to non-dimensionalize $1/T$ by multiplying it by an energy scale. The only energy available is $\phi(r)$, and we write

\begin{equation}
\phi(r)\sim \varepsilon\xi^{-y} \sim \varepsilon\rho^{y/d} \ .
\end{equation}
Thus by presenting $\ln\langle \tau\rangle/\tau_0$ vs. $\varepsilon\rho^{y/d}/T$ we should find data collapse as discussed in Ref. \cite{ludovic}. It is well known that the argument remains approximately valid also when the potential is not a pure power law but with an exponent $x$ that needs to be found by collapsing the data \cite{11PGBSD}. In the second panel of Fig. \ref{fig8} we show the data collapse when plotted properly, with the density varying in the range $[0.80,1.00]$. Additionally, in these systems the inter-particle potential possesses only one energy scale, which is nothing but the depth of the inter-particle potential, denoted as $\varepsilon$. This energy scale must determine the typical barrier height that needs to be surmounted in the Arrhenius regime. Accordingly, we expect $E_a$ to be proportional to $\varepsilon$. This expectation is fully supported by the data shown in Fig. \ref{linear} where for two models (binary Lennard-Jones in 2 and 3 dimensions) we have changed $\varepsilon$ and measured $E_a$.

\begin{figure}
\hskip -1.8 cm
\includegraphics[scale = 0.7]{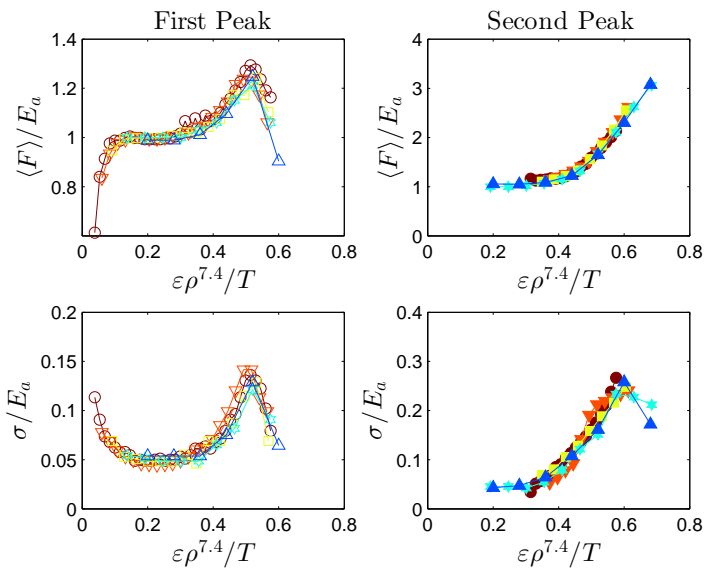}
\caption{Color Online: Data collapse for the normalized energy and dispersion for the same densities as in Fig. \ref{fig8}.}
\label{fig9}
\end{figure}

Our additional contribution to the discussion of density scaling is in pointing out that it is obeyed separately by the two contributions to Eq. (\ref{simple}). In the upper row of Fig. \ref{fig9} we show that $\langle F\rangle$ exhibits a very nice data collapse when data for different densities are re-plotted as required. In the lower row we show the same for $\sigma$. We will use these scaling functions below in providing a model for the observed glassy dynamics.

\section{System size dependence of $T^*$}

\begin{figure}
\hskip -15pt
\includegraphics[scale = 0.7]{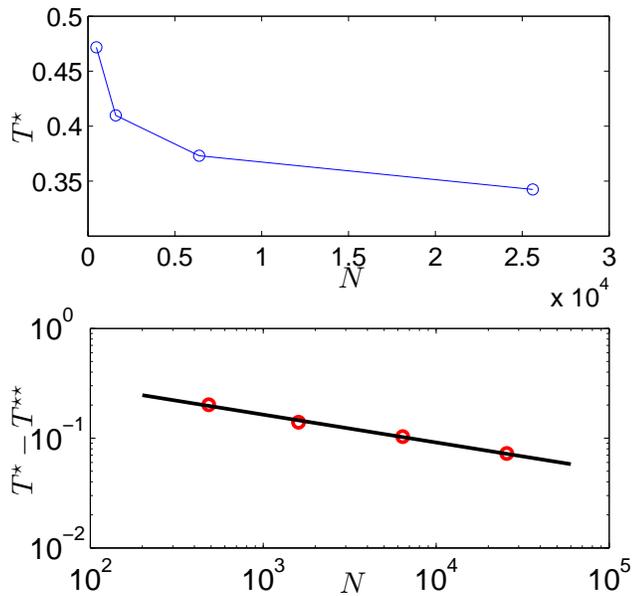}
\caption{Color online: Upper panel: values of $T^\star$ for different system sizes. Lower panel:
 the convergence rate of $T^*$ to the its asymptotic value $T^{**}=0.27$. }
\label{linTstarF}
\end{figure}

An obvious worry about our method of analysis is that the dispersion $\sigma$ depends on the
system size, and therefore the temperature $T^*$ at which we observe the transition to fragility
may depend on the system size. This is of course correct. Nevertheless we discover that the
value of $T^*$ converges to a value $T^{**}$ in the thermodynamic limit. The data is shown for the
binary Lennard-Jones model in Fig. \ref{linTstarF}. From the data in the lower panel one can conclude that
for this model
\begin{equation}
T^*-T^{**} \sim N^{-x} \ , \quad x\approx 0.25 \ .
\end{equation}
We did not perform an exhaustive analysis of the exponents of this law of convergence for the different
models.

\section{cooperative mechanism}

The purpose of this section is to establish one of the main claims of this paper, i.e. that with the
transition to the super-Arrhenius behavior one observes a change from a local to a cooperative relaxation
mechanism.  At temperatures below $T^*$ one expects a cooperative relaxation process to play an important
role; the proposition of this paper is that this is mainly due to the disappearance of the free volume
that was available in the Arrhenius regime (see quantitative model below), requiring now a cooperative motion of many particles to `collect' enough free volume to allow a relaxation step.
Although such a picture is current among some practitioners in the glass community, to the best of
our knowledge it had never been shown explicitly. We use in this section scaling arguments to establish this picture convincingly.

Adding the system size dependence to our typical free energy, we write now

\begin{equation}
F(N,T) = E(T) - T\log[g(N,T)].
\end{equation}
We expect $E(T)$ to be determined by the energy landscape and not to depend strongly on the
system size. Energy barriers are mostly sensitive to local arrangements of particles which are not
system size dependent. On the other hand $\log[g(N,T)]$ is the entropic contribution to the free energy barrier. Our data showed that in the Arrhenius regime the degeneracy factor $g$ was of the order of unity; here we will show that our data strongly supports a cooperative process in the super-Arrhenius regime, mainly due to the system size dependence of the degeneracy factor.

The flip side of having a cooperative mechanism is that there should exist a typical length $\xi(T)$ which measures the degree of cooperativity and is increasing when the temperature decreases.
Consider then a system of $N$ particles in the super-Arrhenius regime, associated with a typical scale $\xi(T)$, and contained in a cubic box of size $L$ at some temperature $T$. As long as the system is small, i.e. $L \ll \xi(T)$, the whole system needs to cooperate in order to relax. Accordingly we expect the degeneracy factor to grow extensively like $g(N,T) \sim N$. When the system is large enough, i.e. when  $L \gg \xi(T)$ we expect to find $g(N,T) \sim \xi^{d}$, not changing with the system size. In other words,
\begin{equation}
g(N,T) \sim \left\{
\begin{array}{ccl}
&& \left(\frac{L}{a}\right)^{ d}, \quad \mbox{for}\quad L \ll \xi(T)\\
&& \left(\frac{\xi(T)}{a}\right)^{ d}, \quad \mbox{for} \quad L \gg \xi(T),
\end{array}\right.
\end{equation}
where $a$ is typical inter particle distance. Accordingly the free energy should assume the form of  a scaling function of $L/\xi(T)$ as
\begin{equation}
F(N,T) = E(T) - T\log\left[A\left(\frac{\xi(T)}{a}\right)^ {d} f\left( \frac{L}{\xi(T)}\right)\right],
\end{equation}
where $A$ is a temperature independent proportionality factor. The scaling function $f(x)$ has the
following asymptotic dependence
\begin{equation}
f(x) = \left\{
\begin{array}{ccl}
&& x^{d} \quad \mbox{for}\quad x \ll 1,\\
&& 1 \quad \mbox{for}\quad x \gg 1.
\end{array}
\right.
\end{equation}
\begin{figure}
\hskip -15pt
\includegraphics[scale = 0.7]{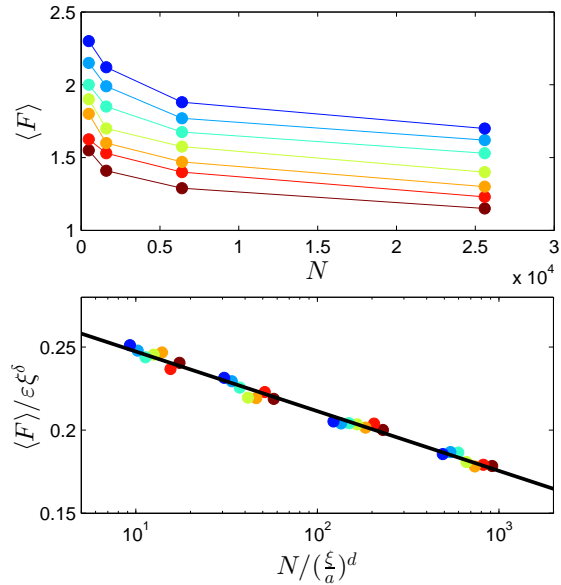}
\caption{(Color online) Upper Panel: The position of the second peak in $P(F,T)$ is plotted as a function of system size for different
temperatures in the range [2.52,3.00]. Lower Panel: Scaling plot (see text for details) to highlight the logarithmic dependence on system size.}
\label{FvsN}
\end{figure}
Considering the small system limit, we conclude that there the free energy barrier should exhibit a logarithmic dependence on the system size. This expectation is supported by the data shown in Fig~\ref{FvsN}.  In the upper panel we show the system size dependence of the free energy value where
the distribution function $P(F,T)$ has a second peak, for different temperatures. With increasing system size these values tend to saturate to an asymptotic value. Note that the dependence on the system size seems to be decreasing with
increasing temperature, indicating the existence of a cooperative length scale. We can not go to high temperatures
in this study because at higher temperature the extraction of the second peak position become difficult as its height goes down drastically, merging with the Arrhenius peak. We focus on data where the
position of the second peak can be estimated accurately. To show the logarithmic dependence on the system size we collapsed the data together  by rescaling the free energy axis by $\xi^\delta$ with
$\delta \simeq 1.2$ and the x axis by $\xi(T)^d$. For simplicity we fitted here a guess function for $\xi(T)$ in the form $\xi(T) \sim 1/T^{1.8}$ \cite{12KLP}. In the lower panel of Fig~\ref{FvsN} one can see the convincing logarithmic dependence of the free energy on the system size for different temperatures.

\section{Interpretation and discussion}
\label{modelsec}
The data shown above indicates two changes in the relaxation mechanism between the pre-Arrhenius and the
Arrhenius regime and then between the Arrhenius and the fragile regime.
We propose that the pre-Arrhenius regime is a fluid regime, in which a well-defined cage is only beginning to form around every particle. It is relatively easy to break this ill-formed cage, and there are many
ways to do it. Accordingly the free energy needed is made of a low value of $E$ and a finite value of $g$
such that when $T$ is increasing the Gaussian peak that we observe marches to the left and becomes broader.

In the Arrhenius regime a well defined cage has formed around each particle, and this cage is a more or less regular cluster of particles aggregating around each center particle. We expect a roughly constant energy $E$ to suffice to break this kind of cage and the width of the distribution of $F$ should be
relatively narrow as discussed below. Towards the end of the Arrhenius regime a second and maybe third
layer of next-nearest neighbors begins to form around a center particle, and therefore it becomes
more and more difficult for the single-particle relaxation to take place. The energy barrier increases but because of the fluidity of the second and third layer also the width of the distribution starts to increase.

Eventually a new, cooperative mechanism must be favored for dynamical relaxation. Note that the apparent mean energy of activation $\langle F \rangle$ {\em increases} when the temperature decreases below the Arrhenius regime. This must mean that the single particle relaxation mechanism that is operative at the Arrhenius regime is no longer available at lower temperatures, since if it were available it would have been selected. In the temperature range where both mechanisms coexists we indeed find the associated $\langle F\rangle$ and $\sigma$ to be similar. We thus offer a model of the observed phenomenology on the the following basis:

(i) In the Arrhenius regime the relaxation process is an elementary
event like a $T_1$ process which is characterized by a relatively sharp free energy cost $F_1$,
\begin{equation}
F_1 \equiv E_a -T\ln g_1  \ , \label{F1}
\end{equation}
where $g_1$ is the number of available inequivalent $T1$ processes. This energy cost $E_a$ is of the order of breaking through a local cage and the process involves only a few particles. Due to the amorphous nature of the super-cooled liquid $E_a$ is a stochastic variable but the dispersion around $E_a$ is relatively small. We denoted the variance in the distribution of $E_a$ as $\sigma^2$ and noted that in this regime $\sigma/E_a\ll 1$.

(ii) The fact that $F_1$ does not change with the temperature in the Arrhenius regime indicates that $g_1$ is of the order of unity.

(iii) The precise reason for the blocking of this elementary relaxation channel when temperature is lowered is not known. We will propose, however, that this blocking is due to the disappearance of available free volume per particle $v_f(T,P)$ \cite{79CG}. In other words, denoting by $v_m$ the minimal average volume that every particle occupies, there must be additional free volume $v_f$ for the $T_1$ process to occur.

Of course, the average volume per particle, which is the inverse of the density, $v(T,P)$, is determined by the equation of state which varies from one material to the other. We will define $T_f$ as the temperature where the free volume disappears, and then $v(T_f,P)=v_m$. For the purposes of the present model we estimate the average volume per particle from the first order Taylor expansion:
\begin{equation}
v(T,P)\approx v(T_f,P) +\left(\frac{\partial v}{\partial T}\right)_P \!\!(T-T_f) \ ,
\end{equation}
where the partial derivative is computed at $T=T_f$.
Accordingly, since $v(T_f,P)=v_m$,
\begin{eqnarray}
v_f(T,P)\! &\!\approx&\!\! \left(\frac{\partial v}{\partial T}\right)_P (T-T_f) \ , \text{provided}~ v>v_m \ , \\
v_f(T,P) &\approx& 0 ~\text{otherwise} \ .
\end{eqnarray}

The probability of having a local free volume $\tilde v_f$ is estimated from maximum entropy considerations as
\begin{equation}
P(\tilde v_f) =\frac{1}{v_f}\exp{-(\tilde v_f/v_f}) \ .
\end{equation}
Consequently the probability that $\tilde v_f>v_m$, which is the probability that a $T1$ process
could occur, is
\begin{equation}
P(\tilde v_f>v_m) = \int_{v_m}^\infty P(\tilde v_f) d\tilde v_f =e^{-v_m/v_f} \ .
\end{equation}
Finally, we write $v_f=v-v_m=\alpha v_m(T-T_f )$, where $\alpha$ is the isobaric coefficient of thermal expansion at $T=T_f$,
\begin{equation}
\alpha=\frac{1}{v_m} \left(\frac{\partial v}{\partial T}\right)_P (T=T_f) \ .
\end{equation}
With this we can write
\begin{equation}
P(\tilde v_f>v_m) =\exp{[-cT_f/(T-T_f)]} \ , \label{probT1}
\end{equation}
where the dimensionless constant $c=1/\alpha T_f$.
Note that we can immediately estimate the probability of seeing the cooperative mechanism (which we
will refer to as the $T_2$ process) as simply $1-\exp{[-cT_f/(T-T_f)]}$. Therefore we can estimate
$T^*$ where the two processes are equiprobable from
\begin{equation}
\!\exp{[-c T_f/(T^*-T_f)]} \approx 1/2 \ ,
\end{equation}
and therefore $ T^*\approx (1+c/\ln2) T_f$.

(iv) In the fragile regime we propose that Eq. (\ref{F1}) retains its form, but now with a temperature dependent $E_2(T)$ and a sizeable degeneracy $g_2(T)$. Our picture is that of a relaxation process that involves more and more cooperative motions of a larger and larger number of particles. This obviously results in an increasing energy barrier $E_2(T)$ which is indeed higher than $E_a$ but selected because the $T_1$ process is no longer available or it requires cooperation from the surroundings to be realized. Also an increase in the number of particles involved must increase $g_2$ simply due to the number of ways that these particles can be chosen. Thus we write in the fragile regime

\begin{equation}
F_2(T) \equiv E_2(T) -T\ln g_2(T)  \ . \label{F2}
\end{equation}

\begin{figure}
\includegraphics[scale = 0.50]{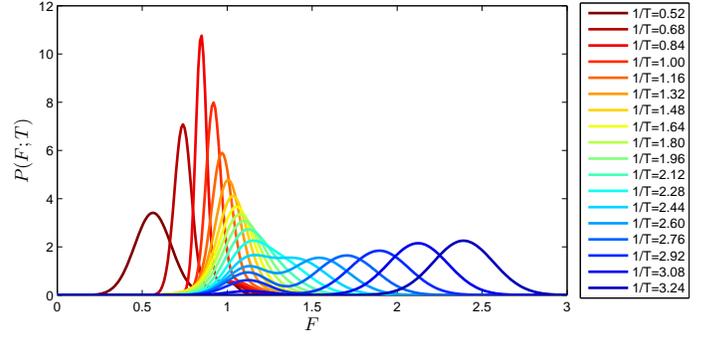}
\caption{Color online. The model prediction for the development of the statistics of $F$ throughout the
temperature range, from pre- to post-Arrhenius behavior. The parameters were chosen to fit the the data of the Lennard-Jones system in 2d, and accordingly the results are in excellent agreement with the numerical
results shown in Fig. \ref{fullT-Lennard}. For much larger system sizes we expect the dispersion $\sigma$ to tend to zero, but this affects neither our model nor the conclusions. }
\label{model1}
\end{figure}

We now consider the emergence of the $T_2$ process as a consequence of the disappearing of the $T_1$ process as is presented for example in Fig. \ref{fullT-Lennard}. We write
\begin{eqnarray}
P(F,T)&=&\frac{W_1(F,T)+W_2(F,T)}{Z(T)} \ , \label{model}\\
Z(T)&=& \int [W_1(F,T)+W_2(F,T)] dF  \ .
\end{eqnarray}
Here $W_1$ and $W_2$ are the weights of the two relaxation channels respectively. In the Arrhenius regime $W_1$ is composed of three factors; remembering that $g_1\approx 1$, we write in that regime
\begin{equation}
W_1(F,T) = e^{[-cT_f/(T-T_f)]}\times \frac{1}{\sqrt{2\pi\sigma^2}}e^{\frac{-(F-E_a )^2}{2\sigma^2}}\times e^{-F/T} \ . \label{W1Ar}
\end{equation}
The first factor is the probability to have a $T_1$ process, Eq. \ref{probT1}. The second factor
is the probability to select a free energy barrier of magnitude $F$. The third factor is the probability to overcome that particular barrier. In the larger temperature regime that includes the pre- and post-Arrhenius regime we generalize Eq.(\ref{W1Ar}) to read
\begin{equation}
W_1(F,T) = e^{[-cT_f/(T-T_f)]}\times \frac{1}{\sqrt{2\pi\sigma_1^2(T)}}e^{\frac{-[F-\langle F_1\rangle(T) ]^2}{2\sigma_1^2(T)}}\times e^{-F/T} \ . \label{W1}
\end{equation}
where $\sigma^2_1(T) \approx \sigma^2 +K(T-T_{\rm Arr})^2$. This fit is supported by the scaling function shown in Fig. \ref{fig9} in the left lower panel, and $T_{\rm Arr}$ is the temperature of the minimum of scaling function.

Similarly,
\begin{equation}
W_2(F,T) = \frac{\{1-e^{[-cT_f/(T-T_f)]}\}}{\sqrt{2\pi\sigma^2_2(T)}}e^{\frac{-[F -\langle F_2\rangle (T) ]^2}{2\sigma^2_2(T)}}\times e^{-F/T}
\end{equation}

In Fig. \ref{model1} we present the prediction of Eq.(\ref{model}) as a function of temperature.
We chose the various parameters to agree with the functions shown in Fig. \ref{fig9} with  $c=0.6$  and $T^*=0.55$. The model indeed
reproduces very well the main observations discussed above.  We see the pre-Arrhenius behavior and its
transition to Arrhenius behavior where expected. We also observe the rapid decline of the Arrehnius gaussian with the
concurrent increase of the fragile gaussian which is marching to the right with increasing dispersion.
We should stress that this simple model is not meant to represent the full complexity of the
observed processes, especially at low temperatures where the cooperative dynamics reigns supreme; the point of this simple model is to stress the fundamental
observation of this paper, i.e. that a compact and well defined relaxation mechanism appears to
operate in the Arrhenius regime, whereas this simple mechanism is blocked once temperature reduces,
giving rise to another mechanism which is presumably cooperative and much more involved. For all that we know one may encounter additional relaxation processes at still lower temperatures which are currently not available to numerical simulations.

\acknowledgments
This work had been supported in part by an advanced ``ideas" grant of the European Research Council, the Israel Science Foundation and the German Israeli Foundation.

\end{document}